\documentclass[runningheads]{llncs}

\usepackage[utf8]{inputenc}
\usepackage{url}
\usepackage{svg}
\usepackage{amsmath,amsfonts}
\usepackage{amssymb}
\usepackage{graphicx}
\usepackage{pgfplots}
\usepackage{tikz}
\usepackage{listings}
\usepackage{color}
\usepackage[dvipsnames]{xcolor}
\definecolor{dkgreen}{rgb}{0,0.6,0}
\definecolor{gray}{rgb}{0.5,0.5,0.5}
\definecolor{mauve}{rgb}{0.58,0,0.82}
\definecolor{grey}{rgb}{0.5,0.5,0.5}
\usepackage[ruled,vlined,linesnumbered,noend]{algorithm2e}
\usepackage[T1]{fontenc}
\usepackage{tabularx}
\usepackage{multirow}
\usepackage{tcolorbox} 
\usepackage{booktabs}
\usepackage{float}
\usepackage{algorithmic}
\algsetup{linenosize=\scriptsize}
\usepackage{setspace}
\usepackage{array}
\usepackage{textcomp}
\usepackage{stfloats}
\usepackage{tcolorbox}
\usepackage{verbatim}
\usepackage{tabularx}
\usepackage{makecell}
\usepackage{bbding}
\usepackage{graphicx}
\usepackage{balance}
\usepackage{threeparttable}
\usepackage{pgfplots}
\usepackage{tikz}
\usepackage{caption}
\usepackage[final]{pdfpages}
\usepackage[commandnameprefix=always, defaultcolor=red]{changes} 
\usepackage[labelformat=simple]{subcaption}

\usepackage{booktabs,multirow,bm,wrapfig,mathtools,dashbox,pifont,soul,enumitem}
\usepackage[commandnameprefix=always, defaultcolor=red]{changes} 

\usepackage[edges]{forest}
\usetikzlibrary{arrows.meta, shadows}
\usetikzlibrary{decorations.pathmorphing}   
\usetikzlibrary{fit}                        
\usetikzlibrary{backgrounds}                
\usetikzlibrary{calc}
\usetikzlibrary{positioning,shadows,arrows}
\usetikzlibrary{shapes,shapes.multipart, shapes.geometric}

\RequirePackage{mdframed}
\renewmdenv[leftmargin=0.4em, rightmargin=0.4em%
innerleftmargin=0.6em,innerrightmargin=0.6em]{quote}

\setlist[itemize]{leftmargin=*}
\setlist[enumerate]{leftmargin=*}
\newlist{steps}{enumerate}{1}
\setlist[steps, 1]{label = \textbf{RQ\arabic*.}}

\usepackage{hyperref}
\usepackage{url}
\usepackage{xcolor,colortbl,color}
\hypersetup{
 colorlinks=true,
 linkcolor=purple,
 citecolor=violet,
 filecolor=magenta,      
 urlcolor=cyan,
}

\newcommand{\eg}{\hbox{\textit{e.g.}}\xspace}

\setlist[itemize]{leftmargin=*}
\setlist[enumerate]{leftmargin=*}

\begin{document}
\title{Large Language Models for Multilingual Code Intelligence: A Survey}

\author{Chao Jiang\inst{1} \and Dugang Liu\inst{1} \and Cheng Wen\inst{2} \and Zhiwu Xu\inst{1}  \and Hua Zheng\inst{3,\star} \and Muhammad Sadiq\inst{4,}\thanks{Corresponding authors: Hua Zheng and Muhammad Sadiq.} \and Jawwad Ahmed Shamsi\inst{5} \and\\ Shengchao Qin\inst{2} \and Zhong Ming\inst{1}}
\institute{College of Computer Science and Software Engineering, Shenzhen University, China \and Guangzhou Institute of Technology, Xidian University, China 
\and Guangzhou University of Software, China \and Shenzhen University of Information Technology \and National University of Computer and Emerging Sciences Karachi, Pakistan
\\
\email{2410104016@mails.szu.edu.cn, zhengh@mail.gzus.edu.cn}
}

\maketitle

\begin{abstract}
\vspace{-15pt}
Large language models have transformed AI-assisted software engineering, but current research remains biased toward high-resource languages such as Python, with weaker performance in languages like \texttt{Rust} and \texttt{OCaml}. Since real-world systems are inherently polyglot, robust multilingual code intelligence is crucial. This survey focuses on two key tasks: multilingual code generation from shared natural-language requirements, and multilingual code translation that preserves semantics across languages. It reviews representative methods, benchmarks, and evaluation metrics, and highlights challenges and opportunities for trustworthy cross-language generalization.

\vspace{-15pt}
\end{abstract}

\vspace{-5pt}
\section{Introduction}\label{sec:intro}
\vspace{-2.5pt}
\textbf{Background.}
The advent of Large Language Models (LLMs)has fundamentally reshaped the landscape of Software Engineering (SE)~\cite{hu2026wen,wen2026when}, particularly in the domain of AI-assisted programming. 
Models such as GPT-5, Deepseek, Qwen3, and Claude-codehave demonstrated remarkable proficiency in understanding natural language intent and generating executable code.
However, current research predominantly exhibits a ``monolingual bias,'' creating a significant disparity in model performance across different programming languages. 
While state-of-the-art LLMs achieve near-human expertise in high-resource languages like \texttt{Python} and \texttt{Java}, their capabilities degrade precipitously when applied to system-level or functional languages such as \texttt{C/C++}, \texttt{Rust}, \texttt{OCaml}, or \texttt{Lua}, a gap clearly evidenced by recent code evaluation leaderboards~\cite{bigcode_leaderboard}.

\textbf{Challenges.}
Multilingual code intelligence is difficult not only because of syntax differences, but also due to the uneven distribution of training and evaluation signals across programming languages.
Public code repositories, tutorials, and Q\&A resources are heavily skewed toward a few popular languages, which biases both pretraining and downstream assessment.
In addition, programming languages embody distinct abstractions and constraints—\textit{e.g.}, ownership and borrowing in \texttt{Rust}, functional idioms and type systems in \texttt{OCaml}, or ecosystem-specific APIs and build tools—which makes cross-language generalization fundamentally harder than ``surface-level'' code formatting.
As a result, a model that performs well on Python-centric tasks may still struggle to produce correct, idiomatic, and buildable implementations in other languages, or may introduce subtle semantic errors that are difficult to detect without rigorous evaluation.

\textbf{Towards Multilingual Code Intelligence.}
Real-world software is inherently \emph{polyglot}. 
Production systems routinely combine multiple programming languages across the stack: a service may expose an HTTP layer in \texttt{Java/Go}, implement performance-critical kernels in \texttt{C/C++}, use \texttt{Rust} for memory safety, rely on \texttt{JavaScript} on the client side, and integrate domain tooling or legacy components in less common languages. 
As software becomes more modular and platform-diverse, multilingual code intelligence becomes essential rather than optional.
In this setting, an LLM is only practically useful if it can 
(i) implement the same intent correctly in different target languages, 
(ii) preserve semantics across cross-language migration, and 
(iii) follow language-specific idioms, libraries, performance requirements, and safety constraints.

\textbf{Two Core Tasks.}
In this survey, we organize multilingual code intelligence around two complementary pillars that reflect common developer workflows:
\begin{enumerate}
\vspace{-5pt}
    \item \textit{Multilingual Code Generation}: generating new implementations from natural-language requirements in one or more target programming languages, which is central to rapid prototyping and deployment across heterogeneous stacks (\eg, generating a \texttt{Python} reference implementation and a \texttt{Rust}/\texttt{C++} production-grade version of the same algorithm), as well as multi-platform delivery. 
    \item \textit{Multilingual Code Translation}: converting an existing program from a source language to a target language while preserving semantics, which is central to modernization, cross-platform porting (\eg, from Windows desktop \texttt{C++} to Android components in \texttt{Java}), and security hardening (\eg, migrating legacy \texttt{C/C++} to memory-safe \texttt{Rust}).
\vspace{-5pt}
\end{enumerate}

\textbf{Scope.}
To keep the survey focused and actionable, we define \emph{multilingual} as involving \emph{at least two distinct programming languages} in the task formulation, model training, or evaluation.
Accordingly, we exclude studies that only generate or evaluate code in a single language, even if the underlying model is nominally multilingual.
Our emphasis is on methods, benchmarks, and analyses that explicitly probe cross-language generalization and semantic preservation.

\vspace{-7.5pt}
\section{Background and Task Formalisms}\label{sec:Background}
\vspace{-5pt}


\vspace{-2.5pt}
\subsection{Formal Notations}
\vspace{-2.5pt}

To discuss the mechanisms of these models, we utilize the formal representations:
\begin{itemize}
\vspace{-5pt}
    \item $NL$ (Natural Language): The set of human-readable languages (\textit{e.g.}, English, Chinese). 
    In this context, $NL$ serves as the high-level intentional input, characterized by its abstraction and potential ambiguity.
    \vspace{1pt}
    \item $PL$ (Programming Language): The set of high-level, executable programming languages (\textit{e.g.}, \texttt{C/C++}, \texttt{Java}, \texttt{Python}, \texttt{JavaScript}, \texttt{Go}, \texttt{Rust}). Unlike $NL$, $PL$ possesses strict syntax and unambiguous semantics.
    \vspace{1pt}
    \item $\{PL_1, PL_2, \dots, PL_n\}$: Denotes a diverse set of distinct programming languages (\textit{e.g.}, $PL_1 = \texttt{Python}$, $PL_2 = \texttt{Rust}$).
    \vspace{1pt}
    \item $PL_{source}$ and $PL_{target}$: Specifically used in translation tasks to denote the origin language and the desired output language, respectively.
\vspace{-5pt}
\end{itemize}

\vspace{-10pt}
\subsection{Multilingual Code Generation}
\vspace{-5pt}

Multilingual code generation, a category of natural language-to-code (NL2Code) tasks, involves synthesizing functional source code from NL descriptions across a diverse range of programming languages.
Formally, this is represented as:\vspace{-7.5pt}
\begin{equation}
\vspace{-10pt}
\begin{aligned}
\mathcal{F}_{gen}: NL \rightarrow \{PL_1, PL_2, \dots, PL_n\}
\end{aligned}
\vspace{2.5pt}
\end{equation}
This task assesses an LLM’s ability to map abstract human intent—the ``what''—to the specific syntactic and idiomatic constraints—the ``how''—of various target languages. 
The primary challenge lies in the model's polyglot competence: it must not only master high-resource languages but also generalize its reasoning to lower-resource or domain-specific languages (DSLs) from a single instruction. 
For example, a model must be able to generate the same algorithmic logic correctly, whether the user requests the output in Python, Java, or Rust, adhering to the unique paradigms and rules of each.
Evaluation typically relies on execution-based benchmarks such as HumanEval-XL~\cite{peng2024humaneval} or McEval~\cite{chai2024mceval}, where the model's performance is measured by its ability to pass a suite of unit tests.

\vspace{-9.5pt}
\subsection{Multilingual Code Translation}
\vspace{-4.5pt}

In contrast to the creative synthesis of generation, Multilingual Code Translation focuses on the automated conversion of existing source code from one programming language to another. 
This is formally represented as:\vspace{-7.5pt}
\begin{equation}
\vspace{-10pt}
\begin{aligned}
\mathcal{F}_{trans}: PL_{source} \rightarrow PL_{target}
\end{aligned}
\vspace{2.5pt}
\end{equation}
The fundamental objective of this task is semantic equivalence: the output code in $PL_{target}$ must produce identical results to the original code in $PL_{source}$ for any given input. 
This is primarily a translating and refactoring task. 
It requires the LLM to possess a deep understanding of divergent programming paradigms.
Translation is critical for high-impact industrial applications. 
A primary use case is software hardening, such as translating legacy \texttt{C/C++} code to \texttt{Rust} to automatically ensure memory safety and eliminate vulnerabilities like buffer overflows. 
Furthermore, translation is essential for cross-platform porting, such as migrating an existing desktop-based \texttt{C++} codebase to the Android mobile system by translating it into \texttt{Java} or \texttt{Kotlin} components. 
Unlike code generation, which is evaluated on problem-solving, translation is assessed through Computational Accuracy (CA) and execution equivalence, utilizing benchmarks like XLCoST~\cite{zhu2022xlcost} and the repository-level RepoTransBench~\cite{wang2024repotransbench}.

\vspace{-9.5pt}
\subsection{Comparative Analysis and Industrial Significance}
\vspace{-4.5pt}

The industrial relevance of these tasks has grown exponentially as software stacks become more fragmented, yet they serve fundamentally different operational goals. 
Multilingual Code Generation acts as a force multiplier for productivity by focusing on the synthesis of new logic to fit different programming environments from a single high-level specification. It is a "top-down" creative process that enables a "write once, generate everywhere" workflow.
In contrast, Multilingual Code Translation addresses the "technical debt" crisis through a "horizontal" transformation, focusing on moving an existing executable program to another expected environment. 
While generation builds from abstract intent, translation preserves concrete execution logic, allowing organizations to modernize aging infrastructures and ensure mobile interoperability without the risks of manual rewriting. 
Together, these complementary capabilities—synthesis and preservation—form the backbone of next-generation AI-assisted software engineering.


\vspace{-10pt}
\section{Methodological Landscape}\label{sec:landscape}
\vspace{-7.5pt}
Research on multilingual code intelligence differs not only in what tasks are addressed, but more fundamentally in how multilingual capability is obtained. 
Existing approaches can be understood along a central axis: whether multilingual generalization is elicited at inference time or learned at training time, and whether semantic robustness is achieved through model-internal representations or external structure and feedback.
From this perspective, we organize the literature into four methodological paradigms, as illustrated in Table~\ref{tab:taxonomy}:
\begin{itemize}
\vspace{-5pt}
    \item \textit{Prompt\,Engineering},\,which\,exploits\,latent\,polyglot\,knowledge\,in\,frozen\,models;
    \vspace{1pt}
    \item \textit{Model Pre-training and Fine-tuning}, which embeds multilingual competence into model parameters;
    \vspace{1pt}
    \item \textit{Multi-Agent Collaborative Frameworks}, which decompose multilingual reasoning into coordinated roles;
    \vspace{1pt}
    \item \textit{Retrieval-Augmented Generation (RAG)}, which grounds multilingual reasoning in external knowledge.
\vspace{-5pt}
\end{itemize}
Rather than treating these paradigms as competing alternatives, we argue that they represent complementary strategies for mitigating different failure modes of multilingual code intelligence.
In addition, these works encompass an extensive array of programming languages—ranging from mainstream languages like Python, Java, and C++ to systems-oriented languages such as Rust and Go, and even specialized languages like Julia and Smalltalk. The models employed represent the current state-of-the-art, reflecting a clear trend toward increasing model capacity and cross-lingual reasoning efficiency.

\begin{table}[t]
\vspace{-5pt}
\scriptsize
\centering
\setstretch{1.05}
\setlength{\tabcolsep}{3.5pt}
\setlength{\abovecaptionskip}{2.5pt}
\setlength{\belowcaptionskip}{-2.5pt}
\caption{Taxonomy of LLMs for Multilingual Generation and Translation.}
\label{tab:taxonomy}
\resizebox{\textwidth}{!}{%
\begin{tabular}{llcclc}
\hline
\rowcolor{blue!05}
\textbf{\small Category} & \textbf{\small Approach} & \textbf{\small Year} & \textbf{\small \begin{tabular}[c]
{@{}l@{}}Programming\\Languages\end{tabular}} & \textbf{\small \begin{tabular}[c]{@{}l@{}}Static Large Language\\Models\end{tabular}} & \textbf{\small \begin{tabular}[c]{@{}l@{}}Task\end{tabular}} 
\\ 
\hline
\rowcolor{gray!05}
& SmartC2Rust~\cite{shiraishi2026smartc2rust} & 2026 & \begin{tabular}[c]{@{}l@{}}C, Rust\end{tabular} & Claude 3.7 Sonnet & Translation \\
& Aljagthami et al.~\cite{aljagthami2025evaluating} & 2025 & \begin{tabular}[c]{@{}l@{}}C++, Java, Python, C\#\end{tabular} & GPT-4o, Gemini 2.0, DeepSeek-V3, Claude 3.7 & Translation \\
\rowcolor{gray!05}
& PSEUDOEVAL~\cite{wu2025isolating} & 2025 & \begin{tabular}[c]{@{}l@{}}Python, C++, Rust\end{tabular} & \begin{tabular}[c]{@{}l@{}}GPT-4o-mini, Qwen-32B, Qwen-7B, Qwen-14B, Phi-4,\\Qwen32Bq4, Llama-8B, Llama-3B, Gemma-9B, Phi-3.5\end{tabular} & Generation \\
& mHumanEval~\cite{raihan2025mhumaneval} & 2025 & \begin{tabular}[c]{@{}l@{}}Python, Java, C++, JavaScript, Ruby, PHP, Go,\\Rust, Fortran, C\#, Lua, TypeScript, Scala, Kotlin, \\R, Dart, Swift, Perl, SQL, Haskell, OCaml, \\Shell, MATLAB, Visual Basic, COBOL\end{tabular} & \begin{tabular}[c]{@{}l@{}}GPT-4o, Claude-3.5-Opus, GPT-3.5, \\DeepSeek-Coder-V2, WizardCoder, Aya\end{tabular} & Generation \\
\rowcolor{gray!05}
& NL in the Middle~\cite{tai2025nl} & 2025 & \begin{tabular}[c]{@{}l@{}}C++, C, Python, Java, Go\end{tabular} & Open GPT4 8X7B, StarCoder, CodeGen & Translation \\
& Ranasinghe et al.~\cite{ranasinghe2025llm} & 2025 & FORTRAN, C++ & \begin{tabular}[c]{@{}l@{}}Mistral Large Instruct 2407, Llama 3.1, \\Llama 3.3, OpenCodeInterpreter\end{tabular} & Translation \\
\rowcolor{gray!05}
& HumanEval-XL~\cite{peng2024humaneval} & 2024 & \begin{tabular}[c]{@{}l@{}}Python, Java, JavaScript, C\#, Go, \\Kotlin, PHP, Perl, Ruby, Swift, \\Scala, TypeScript\end{tabular} & \begin{tabular}[c]{@{}l@{}} CodeT5+, CodeGen2, StarCoder, CodeGeeX, \\GPT-3.5, GPT-4\end{tabular} & Generation \\
& InterTrans~\cite{macedo2024intertrans} & 2024 & \begin{tabular}[c]{@{}l@{}}C++, JavaScript, Java, \\Python, Go, Rust\end{tabular} & Code Llama, Magicoder, StarCoder2 & Translation \\
\rowcolor{gray!05}
\multirow{-16}{*}{\small\textbf{\begin{tabular}[c]{@{}l@{}}Prompt \\Engineering\end{tabular}}} & Buscemi et al.~\cite{buscemi2023comparative} & 2023 & \begin{tabular}[c]{@{}l@{}}C, C++, Go, JavaScript, \\Julia, Perl, Python, R, \\Ruby, Smalltalk \end{tabular} & GPT-3.5 & Generation
 \vspace{5pt} \\ 
\hline
& Giagnorio et al.~\cite{giagnorio2025enhancing} & 2025 & \begin{tabular}[c]{@{}l@{}}Python, Java, Julia, \\Lua, R, Racket\end{tabular} & \begin{tabular}[c]{@{}l@{}}DeepSeek Coder, Code Llama, GitHub Copilot\end{tabular} & Generation \\
\rowcolor{gray!05}
& MCEVAL~\cite{chai2024mceval} & 2024 & \begin{tabular}[c]{@{}l@{}}C++, Java, Python, Go, Rust, \\JavaScript, TypeScript, PHP,\\AWK, Shell, Lisp, Lua, Pascal, \\Fortran, Swift, VimScript\end{tabular} & \begin{tabular}[c]{@{}l@{}}GPT-3.5-Turbo, GPT-4-Turbo, GPT-4o, MCODER,\\CodeQwen1.5, DeepSeekCoder, CodeLlama, Llama3,\\Yi, Phi-3, OCTOCODER, MagiCoder, WizardCoder\end{tabular} & Generation \\
& IRCoder~\cite{paul2024ircoder} & 2024 & \begin{tabular}[c]{@{}l@{}}Python, C++, D, Go, \\Ruby, Rust, Swift\end{tabular} & StarCoderBase, DeepSeekCoder, CodeLlama & Generation \\
\rowcolor{gray!05}
& Magicoder~\cite{wei2023magicoder} & 2023 & \begin{tabular}[c]{@{}l@{}}Java, JavaScript, C++, \\PHP, Swift, Rust, \\TypeScript, Shell, C\#\end{tabular} & CodeLlama-7B-Python, DeepSeek-Coder-Base & Generation \\
\multirow{-10}{*}{\textbf{\small\begin{tabular}[c]{@{}l@{}}Model Pre-\\training and\\Fine-tuning\end{tabular}}} & OCTOPACK~\cite{muennighoff2023octopack} & 2023 & \begin{tabular}[c]{@{}l@{}}Python, JavaScript, Java, \\Go, C++, Rust\end{tabular} & OCTOCODER, OCTOGEEX & Generation 
\vspace{5pt}
\\ 
\hline
\rowcolor{gray!05}
& XL-CoGen~\cite{moumoula2025beyond} & 2025 & \begin{tabular}[c]{@{}l@{}}Python, TypeScript, R, Scala, \\Kotlin, Java, Lua, Julia, PHP, \\Dart, Go Erlang, Haskell, \\Rust, C++, OCaml, Perl\end{tabular} & GPT-4.1-mini, DeepSeek-V3 & Generation \\
& UniPar~\cite{bitan2025unipar} & 2025 & Serial, CUDA, OpenMP & \begin{tabular}[c]{@{}l@{}}GPT-4o-mini, LLaMA-3.3-70B-Instruct\end{tabular} & Translation \\
\rowcolor{gray!05}
& MatchFixAgent~\cite{ibrahimzada2025matchfixagent} & 2025 & \begin{tabular}[c]{@{}l@{}}Go, Rust, Java, Python, \\JavaScript, C\end{tabular}	 & \begin{tabular}[c]{@{}l@{}}Claude 3.7 Sonnet, Claude Code, Codex, \\OpenAI o4-mini\end{tabular} & Translation \\
& Repotransbench~\cite{wang2024repotransbench} & 2024 & \begin{tabular}[c]{@{}l@{}}C, C++, C\#, Java, \\JavaScript, Matlab, Python\end{tabular} & \begin{tabular}[c]{@{}l@{}}Qwen3-235B-A22B, Qwen3-235B-A22B-think, \\DeepSeek-Chat, DeepSeek-Reasoner, Claude-Sonnet-4, \\Gemini-2.5-Flash-Lite, GPT-4.1, o3-mini\end{tabular} & Translation \\
\rowcolor{gray!05}
\multirow{-8}{*}{\textbf{\small\begin{tabular}[c]{@{}l@{}}Multi-Agent \\Collaborative 
\\Frameworks\end{tabular}}} & EVOC2RUST~\cite{wang2025evoc2rust} & 2025 & \begin{tabular}[c]{@{}l@{}}C, Rust\end{tabular} & DeepSeek-V3, Qwen3-32B & Translation
\vspace{5pt}
\\ 
\hline
& Tao et al.~\cite{tao2025retrieval} & 2025 & \begin{tabular}[c]{@{}l@{}}Python, Java, C++, JavaScript, \\C\#, TypeScript\end{tabular} & \begin{tabular}[c]{@{}l@{}}CodeGen, StarCoder, Code Llama, Claude 4/4.5, GPT-5,\\Qwen2.5-Coder, GPT-4o, DeepSeek-Coder, Gemini 2.5 Pro\end{tabular} & Generation \\
\rowcolor{gray!05}
& ARCS~\cite{bhattarai2025arcs} & 2025 & \begin{tabular}[c]{@{}l@{}}Python, Java, C++\end{tabular} & Llama-3 & Generation\\
\multirow{-3}{*}{\textbf{\small\begin{tabular}[c]{@{}l@{}}RAG\end{tabular}}} & Bhattarai et al.~\cite{bhattarai2024enhancing} & 2024 & \begin{tabular}[c]{@{}l@{}}Fortran, C++\end{tabular} &  \begin{tabular}[c]{@{}l@{}}Starcoder, Llama3-70B Instruct, CodeLlama-34B Instruct, \\Granite-34B Code Instruct, Mixtral-8x22B, Codestral, \\Phi-3 3.8B, GPT-3.5, GPT-4o\end{tabular} & Translation
 \vspace{5pt}
\\ 
\hline
\end{tabular}%
}
\vspace{-15pt}
\end{table}

\vspace{-9.5pt}
\subsection{Prompt Engineering}
\vspace{-4.5pt}
Prompt engineering constitutes the most lightweight and widely adopted approach to multilingual code intelligence. 
It assumes a frozen LLM and seeks to elicit multilingual behavior solely through input design, reasoning scaffolds, and in-context examples, without modifying model parameters. 
A significant research focus involves using zero-shot or few-shot prompts to probe performance boundaries and generalization across languages with varying resource levels. 
Benchmarks such as HumanEval-XL~\cite{peng2024humaneval} and mHumanEval~\cite{raihan2025mhumaneval}, along with comparative studies~\cite{buscemi2023comparative}, provide parallel datasets to evaluate how models transition from high-resource languages like Python to lower-resource or specialized ones, often revealing significant performance disparities. 
Expanding on these disparities, Aljagthami et al.~\cite{aljagthami2025evaluating} demonstrate that detailed task specifications and English-based prompts significantly enhance translation quality for frontier models like Claude 3.7 and DeepSeek-V3. 
Their findings also highlight direction-aware asymmetries, where translating into stricter type systems remains more challenging for LLMs than translating into flexible languages like Python.

To mitigate semantic divergence between disparate programming paradigms, recent structural prompting techniques have introduced intermediate representations as semantic anchors for code translation and generation. 
For instance, the NL in the Middle~\cite{tai2025nl} framework utilizes natural language descriptions as a pivot to bridge source and target code, while InterTrans~\cite{macedo2024intertrans} leverages transitive intermediate translations through familiar languages to enhance semantic preservation. 
Beyond surface-level mapping, methods such as PSEUDOEVAL~\cite{wu2025isolating} decouple algorithmic logic from language-specific syntax by guiding models to generate language-agnostic pseudo-code. 
By effectively isolating logical reasoning from coding errors, this paradigm provides a flexible and cost-effective solution for multilingual software engineering, particularly for low-resource languages.


\vspace{-10pt}
\subsection{Model Pre-training and Fine-tuning}
\vspace{-5pt}
In contrast to prompt-centric approaches, parameter-centric methods seek to internalize multilingual code intelligence through targeted pretraining and fine-tuning.
Research in this category primarily focuses on addressing the performance disparity between high-resource and low-resource languages by enriching the training signal through massive data collection or high-quality synthetic generation. Recent efforts focus on large-scale instruction tuning and evaluation across diverse environments. 

To address data scarcity in lower-resource languages, researchers utilize synthetic generation and structural alignment. Magicoder’s~\cite{wei2023magicoder} OSS-Instruct distills knowledge via complex, open-source-derived synthetic pairs, whereas IRCoder~\cite{paul2024ircoder} employs compiler-level Intermediate Representations (IR) as a language-agnostic bridge to preserve semantic logic. Nevertheless, studies like~\cite{giagnorio2025enhancing} emphasize that data quality and diversity remain the bottlenecks, suggesting there is “no silver bullet” for achieving robust proficiency in low-resource languages.


\vspace{-10pt}
\subsection{Multi-Agent Collaborative Frameworks}
\vspace{-5pt}
Multi-agent collaborative frameworks depart from single-pass inference by modeling software engineering as a collaborative, iterative process involving specialized roles.
By simulating iterative software engineering cycles, this approach addresses the inherent limitations of single-shot generation, ensuring better logical consistency and idiomatic correctness in complex multilingual tasks. 
Recent research emphasizes the use of specialized agents to navigate long-tail languages and domain-specific constraints. For instance, XL-CoGen~\cite{moumoula2025beyond} facilitates cross-language semantic alignment for over 20 languages, while frameworks like UniPar~\cite{bitan2025unipar} utilize agentic structures to manage intricate transformations between serial code and parallelized versions such as CUDA or OpenMP.

For repository-level tasks, this paradigm is essential for maintaining semantic equivalence during large-scale migration. RepoTransAgent~\cite{wang2024repotransbench} leverages a ReAct-based multi-agent loop to navigate complex cross-file dependencies and build configurations. Similarly, MatchFixAgent~\cite{ibrahimzada2025matchfixagent} employs an autonomous “validation and repair” loop, guided by static analysis, to eliminate vulnerabilities and functional discrepancies when transitioning legacy C code to memory-safe Rust. Collectively, these advancements move multi-agent collaboration beyond basic code synthesis toward scalable, verifiable software modernization.


\vspace{-10pt}
\subsection{Retrieval-Augmented Generation (RAG)}
\vspace{-5pt}

Retrieval-Augmented Generation (RAG) enhances multilingual code intelligence by injecting external, language-specific knowledge into the inference process.
Research has progressed from basic snippet retrieval toward deep repository-level modeling and cross-language mapping. Tao et al.~\cite{tao2025retrieval} demonstrate that retrieving class definitions and dependencies is vital for code buildability, while Bhattarai et al.~\cite{bhattarai2024enhancing} optimize translation accuracy by selecting semantically similar code pairs from parallel corpora as in-context demonstrations. For low-resource or dynamic settings, frameworks like ARCS~\cite{bhattarai2025arcs} implement agentic “retrieve-synthesize-refine” loops. By iteratively updating context and self-correcting, these methods allow models to master specialized libraries and emerging languages. 
Collectively, RAG provides a project-aware mechanism that grounds static parameters with knowledge, ensuring reliability in large-scale polyglot engineering.


\vspace{-10pt}
\subsection{Comparative Analysis and Synthesis}
\vspace{-5pt}
Across paradigms, a unifying pattern emerges: (1) Prompt engineering improves surface-level alignment but is bounded by the model's pretraining distribution. (2) Pre-training and fine-tuning embed multilingual knowledge but remain constrained by data imbalance, requiring high resources. (3) Multi-agent frameworks provide robustness through decomposition and feedback. (4)  RAG supplies missing domain-specific knowledge at inference time.

Crucially, no single paradigm can fully solve multilingual code intelligence. 
A promising direction lies in hybrid systems that combine parameter-level competence, structured reasoning, external grounding, and iterative validation.

\vspace{-10pt}
\section{Evaluation Practices and Benchmarks}
\vspace{-7.5pt}


Evaluation is a central challenge in multilingual code intelligence. 
Unlike natural language tasks, where approximate semantic similarity may suffice, programming languages demand exactness: minor deviations in logic, typing, or memory behavior can lead to catastrophic failure. This challenge is amplified in multilingual settings, where semantic equivalence must be preserved across divergent programming paradigms~\cite{ma2026integrating,ma2026auto,ma2025Bridging}.

\vspace{-9.5pt}
\subsection{Why Multilingual Code Evaluation Is Fundamentally Hard}
\vspace{-4.5pt}

Multilingual evaluation differs qualitatively from monolingual code assessment for three reasons.
First, semantic equivalence is language-relative. Two programs may be functionally identical yet differ in performance characteristics, memory safety guarantees, or exception behavior depending on the target language. A translation from \texttt{C} to \texttt{Rust} that eliminates undefined behavior is arguably ``better'' than the source—but strictly speaking, it is not behaviorally identical.
Second, evaluation environments are asymmetric. Toolchains, standard libraries, runtime behavior, and testing infrastructure vary widely across languages. Execution-based evaluation in \texttt{Python} is trivial compared to \texttt{Rust} or \texttt{C++}, where compilation, linking, and dependency management introduce additional failure modes unrelated to semantic reasoning.
Third, scale exposes brittleness. While snippet-level correctness can often be validated with unit tests, repository-level multilingual tasks require reasoning over cross-file dependencies, build systems, and configuration artifacts. Most existing benchmarks only partially capture this complexity.
These factors motivate a multi-layered evaluation strategy rather than reliance on a single metric.

\vspace{-9.5pt}
\subsection{Evaluation Metrics}
\vspace{-4.5pt}

\textbf{Execution-based Metrics}. Execution-based metrics evaluate the functional integrity of translated code by dynamically running it within a target environment. \textit{Pass@k} has emerged as the standard automatic metric for evaluating the functional correctness of code generated by Large Language Models (LLMs). It measures the probability that at least one of $k$ distinct generated solutions passes all predefined unit tests for a given problem. Unlike similarity-based metrics, Pass@k directly assesses execution and logical accuracy. \textit{Computational Accuracy (CA)} serves as a fundamental and intuitive metric for evaluating the output correctness of code generated by LLMs. It directly measures the exact match rate between the outputs produced by the generated code for a set of given inputs and the corresponding ground-truth reference outputs. While Pass@k evaluates whether a solution passes all comprehensive unit tests, CA provides a finer-grained, input-output level assessment of result precision.

\textbf{Semantic and textual Metrics}. 
While execution-based metrics verify functional correctness, they often overlook qualitative dimensions such as code readability, maintainability, and security. Semantic and textual metrics are thus employed to evaluate these structural and qualitative attributes. Originally adapted from Natural Language Processing (NLP), metrics such as \textit{BLEU} and its code-specific variant \textit{CodeBLEU} are increasingly marginalized in code generation research. Their primary limitation is a reliance on lexical overlap, which fails to capture the logical brittleness of programming languagesand syntactic diversity. Consequently, these metrics often demonstrate low correlation with actual program correctness. To overcome the constraints of static automated metrics, the \textit{LLM-as-a-Judge} paradigm has emerged as a prominent alternative. This approach utilizes frontier models to evaluate code quality across multi-dimensional criteria, including elegance, logical efficiency, and adherence to complex instructions. While LLM judges offer more comprehensive and expert-level feedback compared to traditional metrics, they face challenges such as inherent model biases and high computational or API costs for large-scale assessments.

\begin{table}[t]
\vspace{-5pt}
\scriptsize
\centering
\setstretch{1.15}
\setlength{\tabcolsep}{7.5pt}
\setlength{\abovecaptionskip}{2.5pt}
\setlength{\belowcaptionskip}{-2.5pt}
\caption{Benchmark Datasets for Multilingual Code Intelligence.}
\label{tab:benchmarks}
\resizebox{\textwidth}{!}{%
\begin{tabular}{llcclc}
\hline
\rowcolor{blue!05}
\textbf{\small \begin{tabular}[c]{@{}l@{}}Benchmark \\Name\end{tabular}} & \textbf{\small Year} & \textbf{\small \begin{tabular}[c]{@{}l@{}}Programming \\Languages\end{tabular}} & \textbf{\small \begin{tabular}[c]{@{}l@{}}Tasks\end{tabular}} 
\\ 
\hline
\rowcolor{grey!05}
mHumanEval~\cite{raihan2025mhumaneval} & 2025 & \begin{tabular}[c]{@{}l@{}}\{Python, Java, C++, JavaScript, Ruby, PHP, Go, Rust, Fortran,\\C\#, Lua, TypeScript, Scala, Kotlin, R, Dart, Swift, Perl, SQL,\\Haskell, OCaml, Shell, MATLAB, Visual Basic, COBOL\}\end{tabular} & <Code Generation> \\
PSEUDOEVAL~\cite{wu2025isolating} & 2025 & \begin{tabular}[c]{@{}l@{}} \;\\ \{Python, C++, Rust\}\\\;\end{tabular} & <Code Generation> \\
\rowcolor{grey!05}
McEval~\cite{chai2024mceval} & 2024 & \begin{tabular}[c]{@{}l@{}}\{AWK, C, C\#, C++, Clojure, CoffeeScript, Common Lisp, Dart, \\Elixir, Emacs Lisp, Erlang, F\#, Fortran, Groovy, Go, HTML, \\Haskell, JSON, Java, JavaScript, Julia, Kotlin, Lua, Markdown,\\PHP, Pascal, Perl, PowerShell, Python, R, Racket, Ruby, \\Rust, Scala, Scheme, Shell, Swift, Tcl, TypeScript, \\VimScript, Visual Basic\}\end{tabular} & \begin{tabular}[c]{@{}c@{}}<Code Generation, \\Code Completion, \\Code Explanation>\end{tabular}\\
HumanEval-XL~\cite{peng2024humaneval} & 2024 & \begin{tabular}[c]{@{}l@{}}\{Python, Java, JavaScript, C\#, Go, Kotlin, \\PHP, Perl, Ruby, Swift, Scala, TypeScript\}\end{tabular} & <Code Generation> \\
\rowcolor{grey!05}
XCODEEVAL~\cite{khan2024xcodeeval} & 2024 &\begin{tabular}[c]{@{}l@{}}\{C, C\#, C++, D, Go, Haskell, Java, JavaScript, Kotlin, \\OCaml, PHP, Pascal, Perl, Python, Ruby, Rust, Scala\}\end{tabular} & \begin{tabular}[c]{@{}c@{}}<Code Generation, Code Translation, \\Code Retrieval>\end{tabular} \\
RepoTransBench~\cite{wang2024repotransbench} & 2024 & \begin{tabular}[c]{@{}l@{}}\{C, C++, C\#, Java, JavaScript, Matlab, Python, Go, Rust\}\end{tabular} & <Code Translation> \\
\rowcolor{grey!05}
Codetransocean~\cite{yan2023codetransocean} & 2023 & \begin{tabular}[c]{@{}l@{}}\{Python, C, C++, C\#, Java, Go, PHP, Visual Basic, Ada, Swift,\\Arturo, AutoHotKey, AWK, BBC Basic, Clojure, Common Lisp,\\Delphi, Elixir, Erlang, Factor, Forth, Fortran, F\#, Groovy, Haskell, \\Icon, Julia, Mathematica, Lua, MATLAB, Nim, OCaml, Pascal,\\ Perl, PowerShell, J, R, Racket, REXX, Ruby, Rust, D, Scala, \\Tcl, COBOL\}\end{tabular} & <Code Translation> \\
G-TransEval~\cite{jiao2023evaluation} & 2023 & \begin{tabular}[c]{@{}c@{}} \,\\\{C++, Java, C\#, Python, JavaScript\}\\,\end{tabular} & <Code Translation> \\
\rowcolor{grey!05}
XLCoST~\cite{zhu2022xlcost} & 2022 & \{C++, Java, Python, C\#, JavaScript, PHP, C\} & \begin{tabular}[c]{@{}c@{}}<Code Generation, Code Translation, \\Code Summarization, Code Search> \end{tabular} \\
CodeXGLUE~\cite{lu2021codexglue} & 2021 & \{Java, C++, C, Python, PHP, JavaScript, Ruby, Go, C\#\} & \begin{tabular}[c]{@{}c@{}}<Code Generation, Code Translation, \\Code Completion, Code Repair, \\Clone Detection, Defect Detection, \\Code Summarization, Code Search>\end{tabular} \\
\rowcolor{grey!05}
CodeNet~\cite{puri2021codenet} & 2021 & \{C++, Python, Java, C, Ruby, C\#\} & \begin{tabular}[c]{@{}c@{}}<Code Similarity, Code Translation, \\Code Classification, Code Repair \\Code Performance Improvement, \\Clone Detection, Code Search, \\Code Completion, Error Prediction>\end{tabular} \\
MBPP~\cite{austin2021program} & 2021 & \{Python\} & \begin{tabular}[c]{@{}c@{}}<Code Generation>\end{tabular}
 \vspace{5pt}
\\ 
\hline
\end{tabular}%
}
\vspace{-15pt}
\end{table}

\vspace{-9.5pt}
\subsection{Benchmarks}
\vspace{-4.5pt}

As shown in Table~\ref{tab:benchmarks}, benchmarks are the diagnostic cornerstone for evaluating multilingual LLMs, following a \textbf{tri-axial evolution}: (1) from single-language to massively multilingual coverage, (2) from lexical matching to execution-based verification, and (3) from isolated snippets to repository-level complexity. 

\textbf{Comprehensive Multi-task Benchmarks}. These benchmarks provide unified frameworks for diverse code intelligence tasks. Foundations were laid by \textsc{CodeNet}~\cite{puri2021codenet}, offering a massive 55-language dataset, and \textsc{CodeXGLUE}~\cite{lu2021codexglue}, which systematized ten tasks including generation, translation, and refinement. Moving beyond static evaluation, \textsc{xCodeEval}~\cite{khan2024xcodeeval} introduced rigorous execution-based verification across 11 languages. Recently, \textsc{McEval}~\cite{chai2024mceval} has extended this scope to include code explanation and code completion, probing the depth of models' multilingual logical understanding. Despite the integration of execution-based verification in modern frameworks, a critical gap remains in addressing the ``long-tail'' challenge posed by domain-specific libraries and legacy systems, which dominate real-world industrial software and hinder the connection between academic evaluation and practical application.

\begin{figure}[t]
\vspace{-2.5pt}
  \begin{minipage}[t]{0.40\linewidth}
    \begin{subfigure}{1\linewidth}
        \centering
        \vspace{-5pt}
        \includegraphics[width=\textwidth]{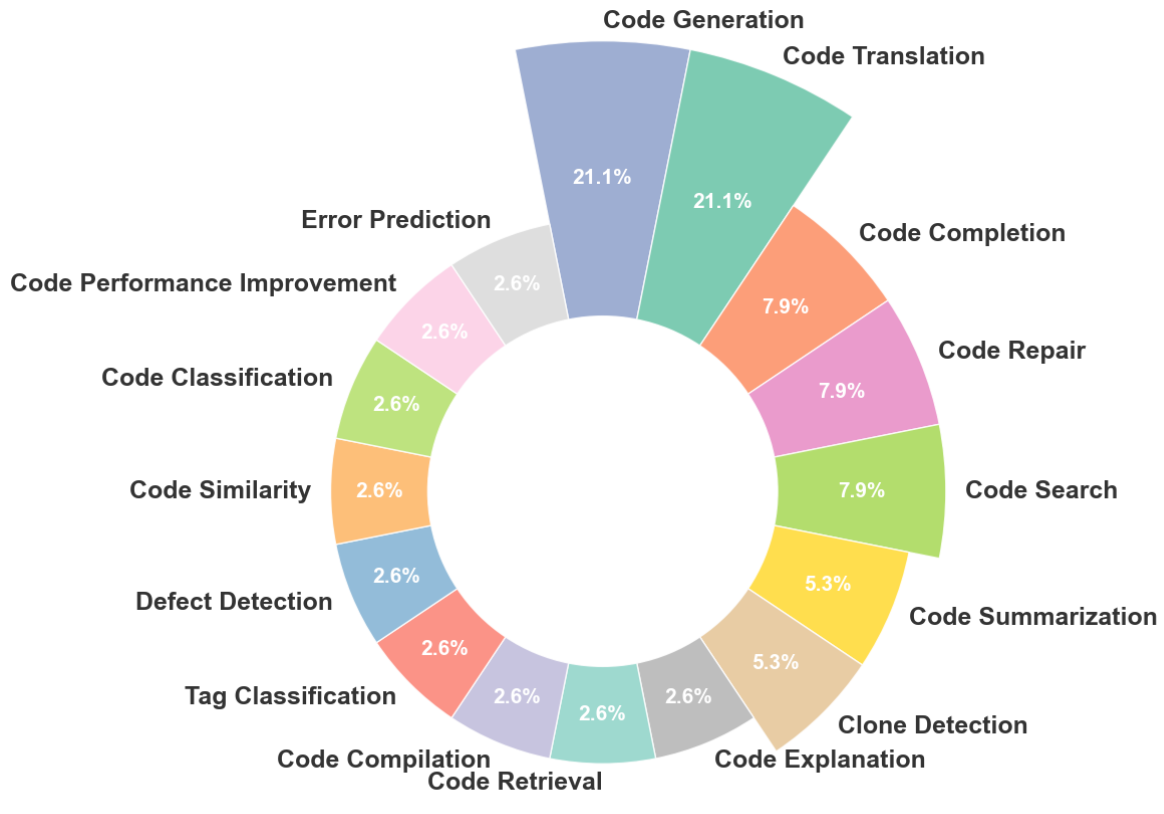}
        \vspace{-0pt}
        \setlength{\abovecaptionskip}{-10pt}
        \setlength{\belowcaptionskip}{0pt}
        \subcaption{\scriptsize Task Distribution in Multilingual Code Intelligence Benchmarks}
        \label{fig:task_distributio}
    \end{subfigure}
  \end{minipage}
  \hspace{0.01\textwidth} 
  \begin{minipage}[t]{0.60\linewidth}
    \begin{subfigure}{1\linewidth}
        \centering
        \vspace{-5pt}
        \includegraphics[width=\textwidth]{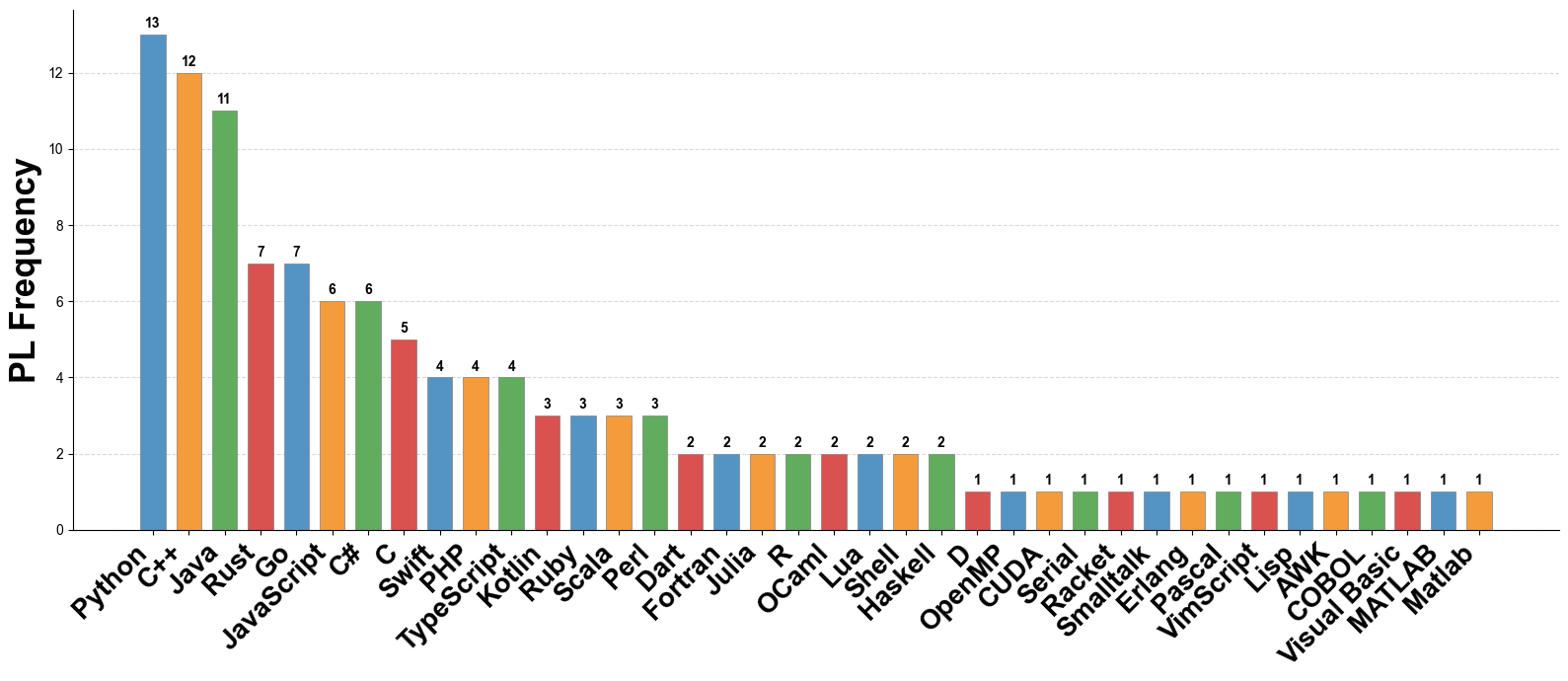}
        \vspace{-0pt}
        \setlength{\abovecaptionskip}{-10pt}
        \setlength{\belowcaptionskip}{0pt}
        \caption{\scriptsize Programming Language Frequency in Surveyed Research}
        \label{fig:pl_frequency}
    \end{subfigure}
  \end{minipage}
  \setlength{\abovecaptionskip}{-7.5pt}
  \setlength{\belowcaptionskip}{-17.5pt}
  \caption{Statistical Data on the Collected Benchmarks}
  \label{fig:datasetstatistics}
  \vspace{-5pt}
\end{figure}


\textbf{Multilingual Code Generation Benchmarks}. This category evaluates the mapping from natural language (NL) instructions to executable code. While \textsc{MBPP}~\cite{austin2021program} initially focused on Python synthesis, \textsc{HumanEval-XL}~\cite{peng2024humaneval} expanded this paradigm to establish a massively multilingual bridge between 23 NLs and 12 PLs. Pushing these boundaries further, \textsc{mHumanEval}~\cite{raihan2025mhumaneval} supports over 200 NLs, providing a globalized lens to assess the cross-lingual instruction-following and code synthesis capabilities of frontier models across low-resource linguistic contexts. Most recently, \textsc{PSEUDOEVAL}~\cite{wu2025isolating} shifts the focus from end-to-end synthesis to component-level isolation by providing solution logic in the form of pseudocode as input. By decoupling problem-solving from language-specific implementation across Python, C++, and Rust, it reveals that while the primary bottleneck for Python lies in logical reasoning, implementation remains a critical barrier for languages with stricter syntax or lower resource availability, such as Rust. Collectively, these advancements highlight a nuanced landscape where the critical barrier for LLMs is no longer solely semantic understanding or cross-lingual transfer, but increasingly the mastery of precise, low-level implementation details within diverse and constrained programming environments.

\textbf{Multilingual Code Translation Benchmarks}. Code-to-code translation benchmarks measure a model's precision in semantic alignment across PLs. XLCoST~\cite{zhu2022xlcost} provided the first fine-grained parallel corpora for this purpose. Subsequently, G-TransEval~\cite{jiao2023evaluation} established a taxonomy to evaluate translation across different complexity levels, while CodeTransOcean~\cite{yan2023codetransocean} significantly increased language coverage to 45 PLs. Recent advancements emphasize real-world engineering complexity: RepoTransBench~\cite{wang2024repotransbench} shifts the focus from isolated snippets to repository-level translation involving cross-file dependencies. 
This evolution mirrors a broader paradigm shift observed across code intelligence benchmarks—from lexical representation and syntactic correctness to functional verification and, increasingly, to system-level and contextual understanding. 
While earlier translation benchmarks tested semantic precision in controlled environments, the latest frameworks probe a model's ability to navigate the architectural and contextual complexities that define practical software migration and maintenance, closing the gap between academic evaluation and industrial applicability.



\vspace{-10pt}
\section{Conclusions}\label{sec:conclu}
\vspace{-7.5pt}
This survey examined progress in multilingual code intelligence, especially code generation and translation across programming languages. Although LLMs have advanced rapidly, they still face monolingual bias, weak cross-paradigm transfer, fragmented evaluation, and limited trustworthiness. Overcoming these issues is key to building reliable polyglot coding systems.


\vspace{-10pt}
\section*{Acknowledgements}
\vspace{-7.5pt}
This work was supported in part by the Basic Research Foundation of Shenzhen City (No. JCYJ20250604184202003), the National Natural Science Foundation of China (Nos. 62302375, 62472339), the China Postdoctoral Science Foundation funded project (No. 2023M723736), the Department of Education of Guangdong Province Foundation (No. 2025KTSCX216), and Guangzhou University of Software Foundation (No. KY202412).
\vspace{-5pt}

\bibliographystyle{unsrt} 
\bibliography{Tex/reference}

\end{document}